\documentstyle[12pt]{article}

\font\twlgot =eufm10 scaled \magstep1
\font\egtgot =eufm8
\font\sevgot =eufm7
\font\twlmsb =msbm10 scaled \magstep1
\font\egtmsb =msbm8
\font\sevmsb =msbm7

\newfam\gotfam

\textfont\gotfam\twlgot
\scriptfont\gotfam\egtgot
\scriptscriptfont\gotfam\sevgot

\newfam\msbfam
\textfont\msbfam\twlmsb
\scriptfont\msbfam\egtmsb
\scriptscriptfont\msbfam\sevmsb
\def\Bbb{\protect\pBbb}
\def\pBbb{\relax\ifmmode\expandafter\Bb\else\typeout{You cann't use
Bbb in text mode}\fi}
\def\Bb #1{{\fam\msbfam\relax#1}}

\def\thebibliography#1{\section*{References}\list
  {[\arabic{enumi}]}{\settowidth\labelwidth{#1}\leftmargin\labelwidth
    \advance\leftmargin\labelsep
    \usecounter{enumi}}
    \def\newblock{\hskip .11em plus .33em minus .07em}
    \sloppy\clubpenalty4000\widowpenalty4000
    \sfcode`\.=1000\relax}

\def\op#1{\mathop{\fam0 #1}\limits}

\newcommand{\id}{{\rm Id\,}}

\newcommand{\beq}{\begin{equation}}
\newcommand{\eeq}{\end{equation}}
\newcommand{\ben}{\begin{eqnarray}}
\newcommand{\een}{\end{eqnarray}}
\newcommand{\be}{\begin{eqnarray*}}
\newcommand{\ee}{\end{eqnarray*}}
\newcommand{\bea}{\begin{eqalph}}
\newcommand{\eea}{\end{eqalph}}

\newcommand{\cA}{{\cal A}}

\newcommand{\cD}{{\cal D}}

\newcommand{\cH}{{\cal H}}

\newcommand{\bL}{{\bf L}}

\newcommand{\al}{\alpha}

\newcommand{\bt}{\beta}

\newcommand{\la}{\lambda}

\newcommand{\Om}{\Omega}
\newcommand{\m}{\mu}
\newcommand{\n}{\nu}
\newcommand{\g}{\gamma}

\newcommand{\vt}{\vartheta}

\newcommand{\lng}{\langle}
\newcommand{\rng}{\rangle}

\newcommand{\w}{\wedge}
\newcommand{\wt}{\widetilde}
\newcommand{\wh}{\widehat}
\newcommand{\ol}{\overline}
\newcommand{\dr}{\partial}

\newcommand{\ot}{\otimes}

\newcounter{eqalph}
\newcounter{equationa}
\newcounter{theorem}
\newcounter{remark}
\newcounter{example}
\newcounter{proposition}
\newcounter{lemma}
\newcounter{corollary}
\newcounter{definition}

\newenvironment{eqalph}{\stepcounter{equation}
\setcounter{equationa}{\value{equation}}
\setcounter{equation}{0}

\begin{eqnarray}}{\end{eqnarray}\setcounter{equation}{\value{equationa}}}

\def\theremark{\arabic{remark}}

\def\thedefinition{\arabic{definition}}

\newenvironment{rem}{\refstepcounter{remark}\bigskip\noindent{\it
Remark \theremark.}}{\medskip}
\newenvironment{ex}{\refstepcounter{remark}\bigskip\noindent{\it
Example \theremark.}}{\medskip}

\newcommand{\mar}[1]{}

\begin{document}
\hbox{}

{\parindent=0pt

{\large\bf Geometric quantization of relativistic
Hamiltonian mechanics}
\bigskip

{\sc Gennadi  Sardanashvily}\footnote{E-mail address:
sard@grav.phys.msu.su; URL: http://webcenter.ru/$\sim$sardan/}
\bigskip

\begin{small}
Department of Theoretical Physics, Physics Faculty, Moscow State
University, 117234 Moscow, Russia

\bigskip



{\bf Abstract.}
A relativistic Hamiltonian mechanical system is seen as 
a conservative Dirac constraint
system on the cotangent bundle of a pseudo-Riemannian
manifold. We provide geometric 
quantization of this cotangent bundle where the quantum constraint
serves as a relativistic quantum equation.   
 
\end{small}
}


\bigskip
\bigskip

We are based on the fact that both
relativistic and  non-relativistic mechanical 
systems on a configuration space $Q$ can be seen as conservative 
Dirac constraint systems on the cotangent bundle $T^*Q$ of $Q$, but
occupy its different subbundles. Therefore, one can follow suit 
of geometric quantization of non-relativistic time-dependent mechanics
in order to quantize relativistic mechanics.

Recall that, given a symplectic manifold $(Z,\Om)$ and a
Hamiltonian $H$ on $Z$, a Dirac constraint system on a closed
imbedded submanifold $i_N:N\to Z$ of $Z$ is defined as a 
Hamiltonian system on $N$ provided with the pull-back presymplectic
form $\Om_N=i^*_N\Om$ and the pull-back Hamiltonian $i^*_NH$ 
\cite{got,book98,mun}. Its solution is a vector field $\g$ on $N$ which
fulfils the equation
\be
\g\rfloor \Om_N +i^*_NdH=0.
\ee
Let $N$ be coisotropic. Then a solution exists if the Poisson bracket
$\{H,f\}$ vanishes on $N$ whenever $f$ is a function
vanishing on $N$. It is the Hamiltonian vector field of $H$ on $Z$ 
restricted to $N$.

A configuration space of non-relativistic time-dependent mechanics
(henceforth NRM) of
$m$ degrees of freedom is an
$(m+1)$-dimensional smooth fibre
bundle $Q\to\Bbb R$ over the time axis $\Bbb R$  \cite{book98,sard98}. 
It is coordinated by
$(q^\la)=(q^0,q^i)$, where $q^0$ is the standard Cartesian
coordinate on $\Bbb R$.
Let $T^*Q$ be the cotangent
bundle of $Q$ equipped with the induced coordinates
$(q^\la,p_\la=\dot q_\la)$ with respect to
the holonomic coframes $\{dq^\la\}$.
Provided with the canonical symplectic form
\mar{q25}\beq
\Om=dp_\la\w dq^\la, \label{q25}
\eeq
the cotangent bundle $T^*Q$
plays the role of a homogeneous momentum phase space
of NRM. Its momentum phase space
is the vertical cotangent bundle $V^*Q$ of $Q\to \Bbb R$ coordinated by
$(q^\la,q^i)$. A Hamiltonian
$\cH$ of NRM is defined as a section
$p_0=-\cH$ of the fibre bundle $T^*Q\to V^*Q$. Then
the momentum phase space of NRM can be identified with the
image $N$ of $\cH$ in $T^*Q$ which is the one-codimensional 
(consequently, coisotropic) imbedded
submanifold given by the constraint
\be
\cH_T=p_0 +\cH(q^\la,p_k)=0. 
\ee
Furthermore, a solution of a non-relativistic Hamiltonian 
system with a Hamiltonian $\cH$ is the restriction $\g$ to $N\cong
V^*Q$ of the Hamiltonian vector field of $\cH_T$ on $T^*Q$.
It obeys the equation
$\g\rfloor \Om_N=0$ \cite{book98,sard98}. Moreover, one can show that
geometric quantization of $V^*Q$ is equivalent to geometric quantization 
of the cotangent bundle
$T^*Q$ where the quantum constraint
$\wh \cH_T\psi=0$
on sections $\psi$ of the quantum bundle serves as the
Schr\"odinger equation \cite{jmp021,jmp022}. This quantization 
is a variant of quantization
of presymplectic manifolds via coisotropic imbeddings \cite{gotsn}.

A configuration space of relativistic mechanics (henceforth RM) is an
oriented pseudo-Riemannian 
manifold $(Q,g)$, coordinated by $(q^\la)$. Its momentum phase space is
the cotangent bundle $T^*Q$ provided with the symplectic form
$\Om$ (\ref{q25}). 
Note that one also considers another symplectic form
$\Om + F$ where $F$ is the
strength of an electromagnetic field \cite{sni}. 
A relativistic
Hamiltonian is defined as a smooth real function
$H$  
on $T^*Q$ \cite{book98,rov91,sard98}. Then a relativistic Hamiltonian
system is described as a Dirac constraint system on
the subspace $N$ of $T^*Q$ given by the equation
\mar{qq90'}\beq
H_T=g_{\m\nu}\dr^\m H\dr^\nu H-1=0. \label{qq90'}
\eeq
Similarly to geometric quantization of NRM,
we provide geometric quantization of the cotangent bundle $T^*Q$ and
characterize a quantum relativistic Hamiltonian system by the
quantum constraint
\mar{qq92'}\beq
\wh H_T\psi=0. \label{qq92'}
\eeq
We choose 
the vertical polarization on $T^*Q$ spanned by the
tangent vectors $\dr^\la$. 
The corresponding quantum algebra $\cA\subset C^\infty(T^*Q)$ consists of
affine functions of momenta 
\mar{ci13}\beq
f=a^\la(q^\m)p_\la +b(q^\m) \label{ci13}
\eeq
on $T^*Q$. 
They are represented by the Schr\"odinger operators
\mar{q1}\beq
\wh f=-ia^\la\dr_\la -\frac{i}{2}\dr_\la a^\la
-\frac{i}{4}a^\la\dr_\la\ln(-g) +b, \qquad g=\det (g_{\al\bt}) \label{q1}
\eeq
in the space $\Bbb C^\infty(Q)$ of smooth complex functions on $Q$.

Note that the function $H_T$ (\ref{qq90'}) need not belong to the quantum
algebra $\cA$. Nevertheless,
one can show that, if $H_T$ is a polynomial of momenta of degree $k$,
it can be represented as a finite composition 
\mar{q3}\beq
H_T=\op\sum_i f_{1i}\cdots f_{ki} \label{q3}
\eeq
of products of affine functions (\ref{ci13}), i.e.,
as an element of the enveloping algebra $\ol\cA$ of the Lie algebra
$\cA$ \cite{jmp021}. Then it is quantized 
\mar{q4}\beq
H_T\mapsto \wh H_T=\op\sum_i \wh f_{1i}\cdots \wh f_{ki} \label{q4}
\eeq
as an element of $\ol\cA$.
However, the representation (\ref{q3}) and, consequently, the
quantization (\ref{q4}) fail to be unique.  

Let us provide the above mentioned formulation of classical RM as a
constraint autonomous mechanics on  
a pseudo-Riemannian manifold $(Q,g)$ \cite{giach99,book98,jmp00}. 
Note that it need not be a space-time manifold. 

The space of relativistic velocities of RM on $Q$ is the
the tangent bundle $TQ$ of $Q$ equipped
with the induced coordinates $(q^\la,\dot q^\la)$ with respect to the
holonomic frames $\{\dr_\la\}$. Relativistic motion is located in
the subbundle $W_g$ of hyperboloids 
\mar{z931}\beq
g_{\m\nu}(q)\dot q^\m\dot q^\nu-1=0
\label{z931}
\eeq
of $TQ$. It is described by a second
order dynamic equation 
\mar{q26}\beq
\ddot q^\la=\Xi^\la(q^\m,\dot q^\m) \label{q26}
\eeq
on $Q$ which preserves the subbundle (\ref{z931}), i.e.,
\be
(\dot q^\la\dr_\la +\Xi^\la\dot\dr_\la)(g_{\m\nu}\dot q^\m\dot
q^\nu-1)=0, \qquad \dot\dr_\la=\dr/\dr\dot q^\la. 
\ee
This condition holds if the right-hand side of the equation (\ref{q26})
takes the form
\be
\Xi^\la=\{_\m{}^\la{}_\n\}\dot q^\m\dot q^\n + F^\la, 
\ee
where $\{_\m{}^\la{}_\n\}$ are  Cristoffel symbols of a metric $g$,
while $F^\la$ obey the relation 
$g_{\m\n}F^\m\dot q^\n=0$.
In particular, if the dynamic equation (\ref{q26}) is a geodesic
equation 
\be
\ddot q^\la = K^\la_\m\dot q^\m 
\ee
with respect to a (non-linear) connection 
\be
K=dq^\la\ot(\dr_\la + K^\m_\la\dot\dr_\m) 
\ee
on the tangent bundle $TQ\to Q$, this connections splits into the sum
\mar{q30}\beq
K_\m^\la=\{\m{}^\la{}_\n\}\dot q^\n +F_\m^\la \label{q30}
\eeq
of the Levi--Civita connection of $g$ and a soldering form 
\be
F=g^{\la\n}F_{\m\n}dq^\m\ot\dot\dr_\la, \qquad F_{\m\n}=-F_{\n\m}.
\ee

As was mentioned above, the momentum phase space of RM on $Q$ is the
cotangent bundle $T^*Q$ 
provided with the symplectic form $\Om$ (\ref{q25}). 
Let $H$ be a smooth real function on $T^*Q$ such that the morphism
\mar{gm616}\beq
\wt H: T^*Q\to TQ, \qquad \dot q^\m=\dr^\m H \label{gm616}
\eeq
is a bundle isomorphism. Then the inverse image $N=\wt H^{-1}(W_g)$ 
of the subbundle of hyperboloids $W_g$ (\ref{z931}) is a
one-codimensional (consequently, coisotropic) closed imbedded subbundle
of $T^*Q$ given by the constraint $H_T=0$ (\ref{qq90'}). We say that
$H$ is a relativistic Hamiltonian if the Poisson bracket
$\{H,H_T\}$ vanishes on $N$. This means that the Hamiltonian vector
field 
\mar{rq11}\beq
\g=\dr^\la H\dr_\la -\dr_\la H\dr^\la \label{rq11}
\eeq
of $H$ preserves the constraint $N$ and, restricted to $N$, it obeys
the Hamilton equation
\mar{gm610}\beq
\g\rfloor \Om_N +i_N^* d H=0 \label{gm610} 
\eeq
of a Dirac constraint system on $N$ with a Hamiltonian $H$.

The morphism (\ref{gm616}) sends the vector field $\g$ (\ref{rq11})
onto the vector field
\be
\g_T=\dot q^\la\dr_\la + (\dr^\m H\dr^\la\dr_\m H-\dr_\m H\dr^\la\dr^\m
H)\dot\dr_\la
\ee
on $TQ$. This vector field defines the second order dynamic equation 
\mar{q35}\beq
\ddot q^\la=\dr^\m H\dr^\la\dr_\m H-\dr_\m H\dr^\la\dr^\m H \label{q35}
\eeq
on $Q$ which preserves the subbundle of hyperboloids (\ref{z931}).

\begin{ex}
The following is a basic example of relativistic Hamiltonian systems.
Put
\be
H= \frac1{2m}g^{\m\n}(p_\m-b_\m)( p_\nu-b_\nu), 
\ee
where $m$ is a constant and $b_\m dq^\m$ is a covector field on $Q$. Then
$H_T=2m^{-1}H-1$ and $\{H,H_T\}=0$. The constraint
$H_T=0$ defines a closed imbedded one-codimensional subbundle $N$ of $T^*Q$.
The Hamilton equation
(\ref{gm610}) takes the form $\g\rfloor\Om_N=0$. Its solution
(\ref{rq11}) reads
\be
&& \dot q^\al=\frac1{m}g^{\al\nu}(p_\nu -b_\nu), \\
&& \dot p_\al=-\frac1{2m}\dr_\al g^{\m\nu}(p_\m-b_\m)( p_\nu-b_\nu)
+\frac1m g^{\m\nu}(p_\m-b_\m)\dr_\al b_\nu. 
\ee
The corresponding second order dynamic equation (\ref{q35}) on $Q$ is
\mar{q12}\ben
&& \ddot q^\la=\{_\m{}^\la{}_\nu\}\dot q^\m\dot q^\nu - \frac1m
g^{\la\nu}F_{\m\nu} \dot q^\mu, \label{q12}\\
&& \{_\m{}^\la{}_\nu\}=-\frac12g^{\la\bt}(\dr_\m g_{\bt\nu}+\dr_\nu
g_{\bt\m} -\dr_\bt g_{\m\nu}), \qquad F_{\m\nu}=\dr_\m b_\nu-\dr_\nu b_\m.
\nonumber
\een
It is a geodesic equation with respect to the affine connection
\be
K^\la_\m=\{_\m{}^\la{}_\nu\}\dot q^\nu - \frac1m
g^{\la\nu}F_{\m\nu} 
\ee
of type (\ref{q30}).
For instance, let $g$ be a metric gravitational field and 
let $b_\m=eA_\m$,
where $A_\m$ is an electromagnetic potential whose gauge holds fixed.
Then the equation 
(\ref{q12}) is the well-known equation of motion of a 
relativistic massive charge in the presence of these fields. 
\end{ex}

Turn now to quantization of RM.
We follow the standard geometric quantization of the cotangent bundle 
\cite{blat,sni,wood}.
Because the canonical symplectic form $\Om$ (\ref{q25}) on $T^*Q$ is exact, 
the prequantum bundle
is defined as a trivial complex line bundle $C$ over $T^*Q$.
Note that this bundle need no metaplectic
correction since $T^*X$ is endowed with canonical coordinates 
for the symplectic
form $\Om$. Thus, $C$ is a quantum bundle.
Let its trivialization
\mar{ci3}\beq
C\cong T^*Q \times \Bbb C \label{ci3}
\eeq
hold fixed, and let $(q^\la,p_\la,c)$, $c\in\Bbb C$, 
be the associated bundle coordinates. 
Then one can treat sections of $C$ (\ref{ci3}) 
as smooth complex functions on $T^*Q$.
Note that another trivialization of $C$ leads to an
equivalent quantization of $T^*Q$.

The Kostant--Souriau prequantization formula associates to
each smooth real function $f\in C^\infty(T^*Q)$ on
$T^*Q$ the first order differential operator
\mar{lqq46}\beq
\wh f=-i\nabla_{\vt_f} + f \label{lqq46}
\eeq
on sections of $C$, where $\vt_f=\dr^\la f\dr_\la -\dr_\la f\dr^\la$
is the Hamiltonian vector field of $f$ and
$\nabla$ is the covariant differential with respect to a
suitable $U(1)$-principal connection $A$ on $C$. This connection
preserves the
Hermitian metric $g(c,c')=c\ol c'$ on $C$, and
its curvature form obeys the prequantization
condition $R=i\Om$. For the sake of simplicity, let us assume that $Q$
and, consequently, $T^*Q$ is simply connected. Then the connection $A$
up to gauge transformations is 
\mar{ci14}\beq
A=dp_\la\ot\dr^\la + dq^\la\ot(\dr_\la +icp_\la\dr_c), \label{ci14}
\eeq
and the prequantization operators (\ref{lqq46}) read
\mar{ci4}\beq
\wh f=-i\vt_f +(f-p_\la\dr^\la f). \label{ci4}
\eeq

Let us choose the vertical polarization on $T^*Q$. It is 
the vertical tangent bundle $VT^*Q$ of the fibration $\pi:T^*Q\to Q$.
As was mentioned above, the corresponding quantum algebra
$\cA\subset C^\infty(T^*Q)$
consists of affine functions $f$ (\ref{ci13}) of momenta $p_\la$.
Its representation by operators (\ref{ci4}) is
defined in the
space $E$ of sections $\rho$ of the quantum bundle $C$ of
compact support
which obey the condition $\nabla_\vt\rho=0$ for any vertical
Hamiltonian vector field 
$\vt$ on $T^*Q$. This condition takes the form
\be
\dr_\la f\dr^\la\rho=0, \qquad \forall f\in C^\infty(Q).
\ee
It follows that elements of $E$ are independent of momenta and,
consequently, fail to be compactly supported, unless $\rho=0$.
This is the well-known problem of Schr\"odinger quantization which
is solved as follows \cite{blat,jmp021}.

Let $i_Q:Q\to T^*Q$ be the canonical zero section of the cotangent
bundle $T^*Q$.
Let $C_Q=i^*_QC$ be the pull-back of the bundle $C$ (\ref{ci3})
over $Q$. It is a trivial complex line bundle $C_Q=Q\times\Bbb C$
provided with the pull-back Hermitian metric
$g(c,c')=c\ol c'$ and the pull-back
\be
A_Q = i^*_QA= dq^\la\ot(\dr_\la +icp_\la\dr_c)
\ee
of the connection $A$ (\ref{ci14}) on $C$.
Sections of $C_Q$ are smooth complex functions on
$Q$, but this bundle need metaplectic correction.

Let the cohomology group $H^2(Q;\Bbb Z_2)$ of $Q$ be trivial. Then
a metalinear bundle $\cD$
of complex half-forms on $Q$ is defined.
It admits the canonical lift
of any vector field $\tau$ on $Q$ such that
the corresponding
Lie derivative of its sections reads
\be
\bL_\tau=\tau^\la\dr_\la+\frac12\dr_\la\tau^\la.
\ee
Let us consider the tensor product
$Y=C_Q\ot\cD$ over $Q$. 
Since the Hamiltonian vector fields
\be
\vt_f=a^\la\dr_\la-(p_\m\dr_\la a^\m +\dr_\la b)\dr^\la
\ee
of functions $f$ (\ref{ci13}) are projected onto $Q$, one can
assign to each
element $f$ of the quantum algebra $\cA$ the first order
differential operator
\be
\wh f=(-i\ol\nabla_{\pi\vt_f} +f)\ot\id+\id\ot\bL_{\pi \vt_f}=
-ia^\la\dr_\la-\frac{i}{2}\dr_\la a^\la +b 
\ee
on sections $\rho_Q$ of $Y$. For the sake of simplicity, let us choose
a trivial metalinear bundle $\cD\to Q$ associated to the orientation of $Q$.
Its sections can be written in the form $\rho_Q=(-g)^{1/4}\psi$, where
$\psi$ are smooth complex functions on $Q$. Then the
quantum algebra $\cA$ can be represented by the operators $\wh f$
(\ref{q1}) in the space $\Bbb C^\infty(Q)$ of these functions.
It is easily justified that these operators obey the Dirac condition
\be
[\wh f,\wh f']=-i\wh{\{f,f'\}}.
\ee

\begin{rem}
One usually considers the subspace 
$E_Q\subset \Bbb C^\infty(Q)$ of functions of compact
support. It is a pre-Hilbert space with respect to the non-degenerate
Hermitian form
\be
\lng \psi|\psi'\rng=\op\int_Q
\psi \ol\psi'(-g)^{1/2}d^{m+1}q
\ee
It is readily observed that $\wh f$ (\ref{q1}) are symmetric operators
$\wh f=\wh f^*$ in $E_Q$, i.e., 
$\lng\wh f\psi|\psi'\rng=\lng\psi|\wh f\psi'\rng$. In RM,
the space $E_Q$ however gets no physical meaning.
\end{rem}

As was mentioned above, the function $H_T$ (\ref{qq90'}) need not
belong to the quantum algebra $\cA$, but a polynomial function $H_T$
can be quantized as an element of the enveloping algebra $\ol\cA$ by
operators $\wh H_T$ (\ref{q4}). Then the quantum constraint (\ref{qq92'})
serves as a relativistic quantum equation.

\begin{ex}
Let us consider a massive relativistic charge in Example 1
whose relativistic Hamiltonian is
\be
H=\frac1{2m}g^{\m\n}(p_\m-eA_\m)(p_\n-eA_\n).
\ee
It defines the constraint
\mar{q20}\beq
H_T=\frac1{m^2}g^{\m\n}(p_\m-eA_\m) (p_\n-eA_\n)-1=0. \label{q20}
\eeq
Let us represent the function $H_T$ (\ref{q20}) as the symmetric product
\be
H_T=\frac{(-g)^{-1/4}}{m}\cdot (p_\m-eA_\m)\cdot(-g)^{1/4}\cdot g^{\m\n}\cdot
(-g)^{1/4}\cdot (p_\n-eA_\n)\cdot \frac{(-g)^{-1/4}}{m}-1 
\ee
of affine functions of momenta. It is quantized by the rule (\ref{q4}), 
where 
\be
(-g)^{1/4}\circ \wh\dr_\al\circ (-g)^{-1/4}=-i\dr_\al.
\ee
Then the well-known relativistic quantum equation
\mar{q41}\beq
(-g)^{-1/2}[(\dr_\m-ieA_\m)g^{\m\n}(-g)^{1/2}(\dr_\n-ieA_\n)+m^2]\psi=0.
\label{q41}
\eeq
is reproduced up to the factor $(-g)^{-1/2}$.
\end{ex}


\end{document}